\documentclass[twocolumn,showpacs,preprintnumbers,amsmath,amssymb]{revtex4}

\usepackage{graphicx}
\usepackage{dcolumn}
\usepackage{bm}
\usepackage{color}

\begin{document}

\title{ Rogue waves emerging from the resonant interaction of three waves}

%
%
%
%
%

\author{Fabio Baronio$^1$, Matteo Conforti$^1$, Antonio Degasperis$^2$ and Sara Lombardo$^3$ }
\affiliation{$^1$ Dipartimento di Ingegneria dell'Informazione, Universit\`a di Brescia, Via Branze 38, 25123 Brescia, Italy\\ $^2$ INFN, Dipartimento di Fisica, ``Sapienza'' Universit\`a di Roma, P.le A. Moro 2, 00185 Roma, Italy
 \\ $^3$ Department of Mathematics and Information Sciences, Northumbria University, Newcastle upon Tyne, United Kingdom}

\date{\today}

\begin{abstract}
We introduce a novel family of analytic solutions of the three-wave resonant interaction equations to the purpose of modeling unique events, i.e. ``amplitude peaks'' which are isolated in space and time.  
%
The description of these solutions is likely to be a crucial step in the understanding and forecasting of rogue-waves in a variety of multi-component wave dynamics, from oceanography to optics, from plasma physics to acoustics.  
\end{abstract}

\maketitle

\textit{Introduction.}
The nature of rogue waves, mostly known as oceanic phenomena
responsible for a large number of maritime disasters, has been discussed in the literature for decades \cite{draper65,dean90,muller05,perkin06}. 
A number of various approaches have been suggested to explain the high-impact power of these ``monsters of the deep'' \cite{kharif09}, which appear visibly from nowhere and disappear without a trace. 
Theories may differ depending on the physical conditions where these waves appear \cite{erkin2009,akhmediev2010sp}.
%
%


As a matter of facts, a comprehensive understanding of the protean rogue wave phenomenon is still far from being acheived \cite{kharif09,osborne10}. Indeed these waves
not only appear in oceans  but also in the atmosphere \cite{stenflo2009}, 
in optics \cite{erkin2009}, in plasmas \cite{moslem2011}, in superfluids \cite{ganshin2008}, in Bose-Einstein
condensates \cite{bludov2009} and capillary waves \cite{xia2010}.
Peculiar aspects and common features of the multifaceted manifestations of rogue waves in their different physical realms are a subject of intense scientific debate \cite{akhmediev2010sp}. New studies of rogue waves in any of these 
disciplines contribute to give a global view on a complex process that to a large extent remains unexplored \cite{onorato13}.

Nonlinear dynamics is one of the theoretical framework that has been successful in predicting the basic features of rogue waves \cite{onorato01,garrett09}. A formal prototypical description of a single rogue wave is provided by the so-called Peregrine soliton, a solution of the  focusing nonlinear  Schr\"odinger equation (NLSE)
\cite{peregrine83,shrira09} which features a rational dependence on both space and time coordinates. Such solution describes the growing evolution
of a small, localized perturbation of a plane wave whose peak subsequently gets amplified by a maximal factor 
 $3$ over the background and eventually decays and vanishes. 
After decades of debate \cite{kharif09,akhmediev2010sp}, the Peregrine soliton has been observed 
experimentally only very recently in fiber optics \cite{kibler10}, in water-wave tanks \cite{amin11}, 
and in plasmas \cite{bailung11}. 
%
%
%
Moreover the Peregrine soliton turns out to be just the first of  an infinite hierarchy of higher order rational solitons of the focusing NLSE with a progressively increasing peak amplitude. Again their amplitude over the background is expressed by the ratio of progressively higher degree polynomials and are therefore localized both in space and time \cite{ACA10,amin12}. 
These theoretical findings and experimental observations prove  
that the approach based on fundamental nonlinear models, such as NLSE,  may be fruitful 
and appropriate to rogue waves description.

In a variety of physical contexts, several waves rather than a single one need to be considered, in order to account for important resonant interaction processes. In these circumstances  extreme waves should be described as solutions of coupled systems of equations rather than by the single-wave NLSE model. In this direction the investigation of solutions which are possible candidates as rogue waves has been recently extended to coupled NLSEs \cite{baronio12,onorato12g,liu2013,zhai2013}. This weak resonant interaction of two waves has been shown to cause wave behaviors which could not be detected by the single Peregrine soliton. However, in order to account for the appearance and dynamics of extreme waves in strong resonant processes, the three-wave resonant interaction (TWRI) equations seem to be the fundamental and universal model. Indeed this system describes the propagation and mixing of waves in weakly nonlinear and dispersive 
media. Applications are found in fluid-dynamics (capillary-gravity waves, internal gravity waves, surface and internal waves), 
in optics (parametric amplification, frequency conversion, stimulated Raman and Brillouin scattering), in plasmas
(plasma instability, laser-plasma interactions, radio frequency heating), in acoustics and solid-state physics. 

In this Letter, we introduce, for the first time to our knowledge,  a family of rational multi-component solutions of 
the TWRI equations which describe unique events, i.e. 
``amplitude peaks'' which are distinctive of rogue waves as being much higher than the surrounding background and well isolated in both space and time. These solutions are expected to be crucial in forecasting and explaining extreme waves  in a variety of multi-component resonant processes (f.i. oceanography, optics, plasma physics).

\textit{TWRI equations and Rogue Waves.}\label{sec2}
The system of three coupled partial differential equations we chose here to model 
 the resonant interaction of three waves in $1 + 1$ dimensions reads as follows (in the notation of  \cite{deg06}):
\begin{eqnarray}\label{3wri}
\nonumber E_{1t}+V_1 E_{1z}&=&\phantom{-} E_2^*E_3^*,\\ E_{2t}+V_2
E_{2z}&=&-E_1^*\,E_3^*,\\ \nonumber E_{3t}+V_3
E_{3z}&=&\phantom{-} E_1^*\,E_2^*,
\end{eqnarray}
where each subscript variable stands for partial differentiation. 
$E_n=E_n(z,t)$, $n=1,2,3$, are complex amplitudes, $t$ is the evolution variable 
and $z$ is a second independent variable. The coefficients $V_n$  are the 
velocities of the three waves and we assume the ordering $V_1>V_2>V_3$. 
With no loss of generality, we assume $V_3=0$ by writing these
equations (\ref{3wri}) in the reference frame moving with the same velocity of $E_3$.
The signs of the coupling constants, with the minus sign only in the equation with 
the intermediate velocity $V_2$, correspond to the so-called ``soliton-exchange" case in the terminology of  \cite{KR79}. 

It should be pointed out that the meaning of the complex amplitude 
$E_n$, and of the coordinates $t,z$, 
depends on the particular applicative context (f.i., fluid-dynamics \cite{lamb07}, 
plasma physics \cite{dodin02}, nonlinear optics \cite{baronio10}, acoustics
\cite{burlak00}). 

TWRI equations, like NLSE, possess rational solutions with the property of representing, in each of the 
three waves $E_n$, ``amplitude peaks'' which are isolated in space and time. 
Similarly to the case of the NLSE, these solutions are local deformations of a 
non vanishing background whose modulation instability is discussed in \cite{CBD11}. 
Such solutions can be expressed as:
\begin{subequations}\label{pere}
\begin{equation}\label{E1}
E_1=2 q \delta_1
  \big[1+\frac{3 \sqrt{3} A^* \theta^* A_1}{|A|^2+|A_1|^2+|A_2|^2}  \big] e^{iq( t- \nu_1 z)}
\end{equation}
\begin{equation} \label{E2}
E_2=2 q  \delta_2
 \big[1+\frac{3 \sqrt{3}  A \theta^* A_2^*}{|A|^2+|A_1|^2+|A_2|^2}  \big] e^{iq( t+ \nu_2 z)} 
\end{equation}
\begin{equation} \label{E3}
E_3=2 i q \delta_3
   \big[1+\frac{3 \sqrt{3} \theta^* A_1^* A_2}{|A|^2+|A_1|^2+|A_2|^2}  \big] e^{-iq[2 t+ (\nu_2-\nu_1) z]}
\end{equation}
\end{subequations}
where
\begin{gather*}\label{def}
 \theta=(- \sqrt{3}+i)/2, |\theta|=1,   \\
\delta_1=\sqrt{(V_1-V_2)/V_1}, \delta_2=\sqrt{(V_1-V_2)/V_2},  \\ 
\delta_3=(V_1-V_2)/ \sqrt{V_1 V_2}, \\ 
A=\gamma_1+ \gamma_2 \xi_1 + \gamma_3 (\eta- i \theta^*), \\
A_1=\gamma_1+ \gamma_2 (\xi_1+\theta^*) + \gamma_3 (\eta+ \theta^* \xi_1+ i \sqrt{3} ), \\
A_2=\gamma_1+ \gamma_2 (\xi_1+\theta) + \gamma_3 (\eta+ \theta \xi_1), \\
\xi_1=-2q( t +i \rho_1 z), \ \ \eta=\frac12 \xi_1^2- 2iq \rho_2 z,   \\
\rho_1= \theta/V_1-\theta^*/V_2,  \ \ \ \rho_2= 1/V_1-1/V_2, \\
\nu_1= 2/V_2-1/V_1, \\
\nu_2= 1/V_2-2/V_1.  
%
%
%
\end{gather*}
The above expressions depend on the velocities, $V_1, V_2$, the real ``frequency'' parameter $q$, 
and the complex parameters $\gamma_1, \gamma_2, \gamma_3$. 
Once the structural parameters (i.e. the characteristic velocities $V_1$ and $V_2$) 
are fixed, we are left with four independent parameters $q$, and $\gamma_1, 
\gamma_2, \gamma_3$. 

\begin{figure}[h]
\begin{center}
\includegraphics[width=6cm]{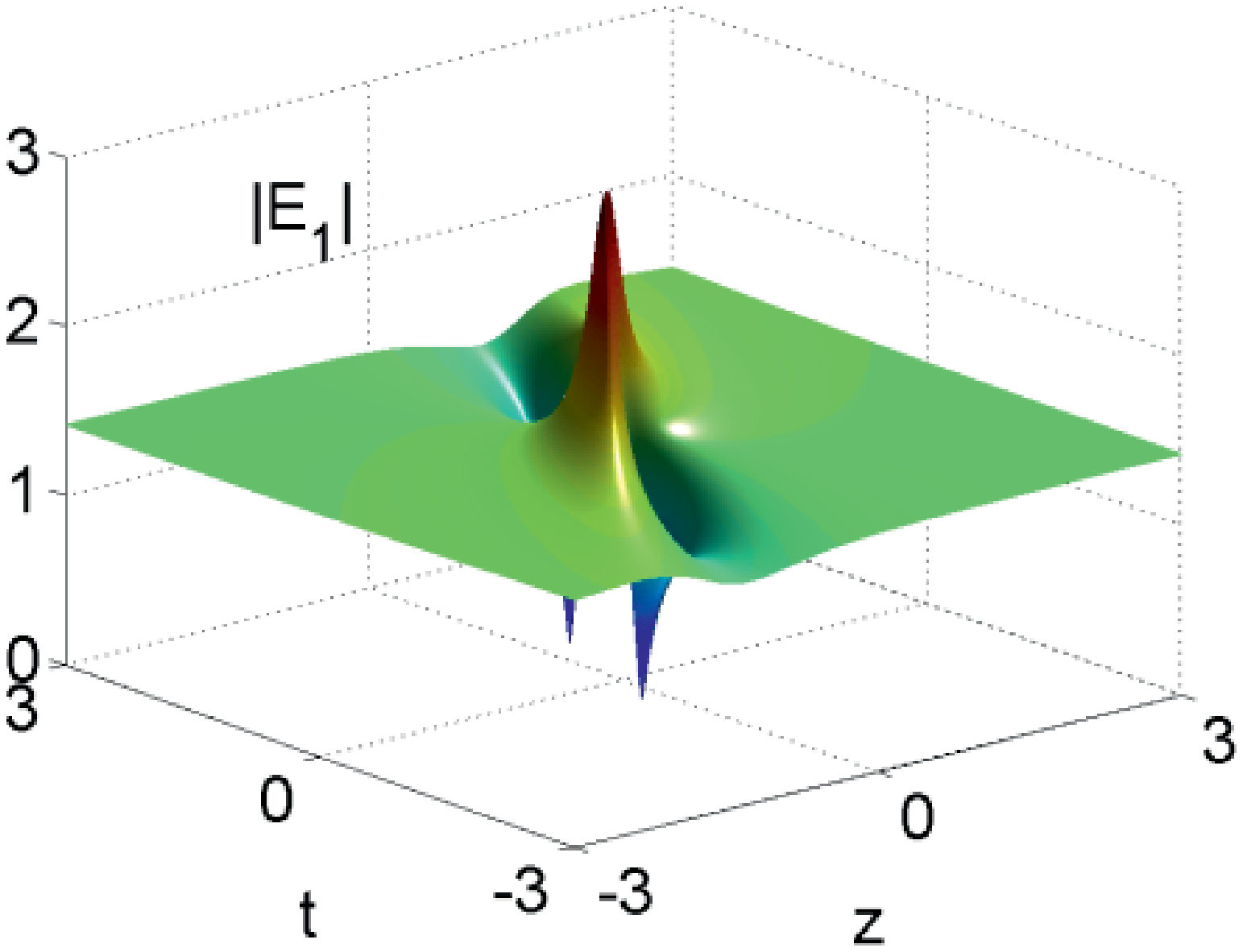}
\includegraphics[width=6cm]{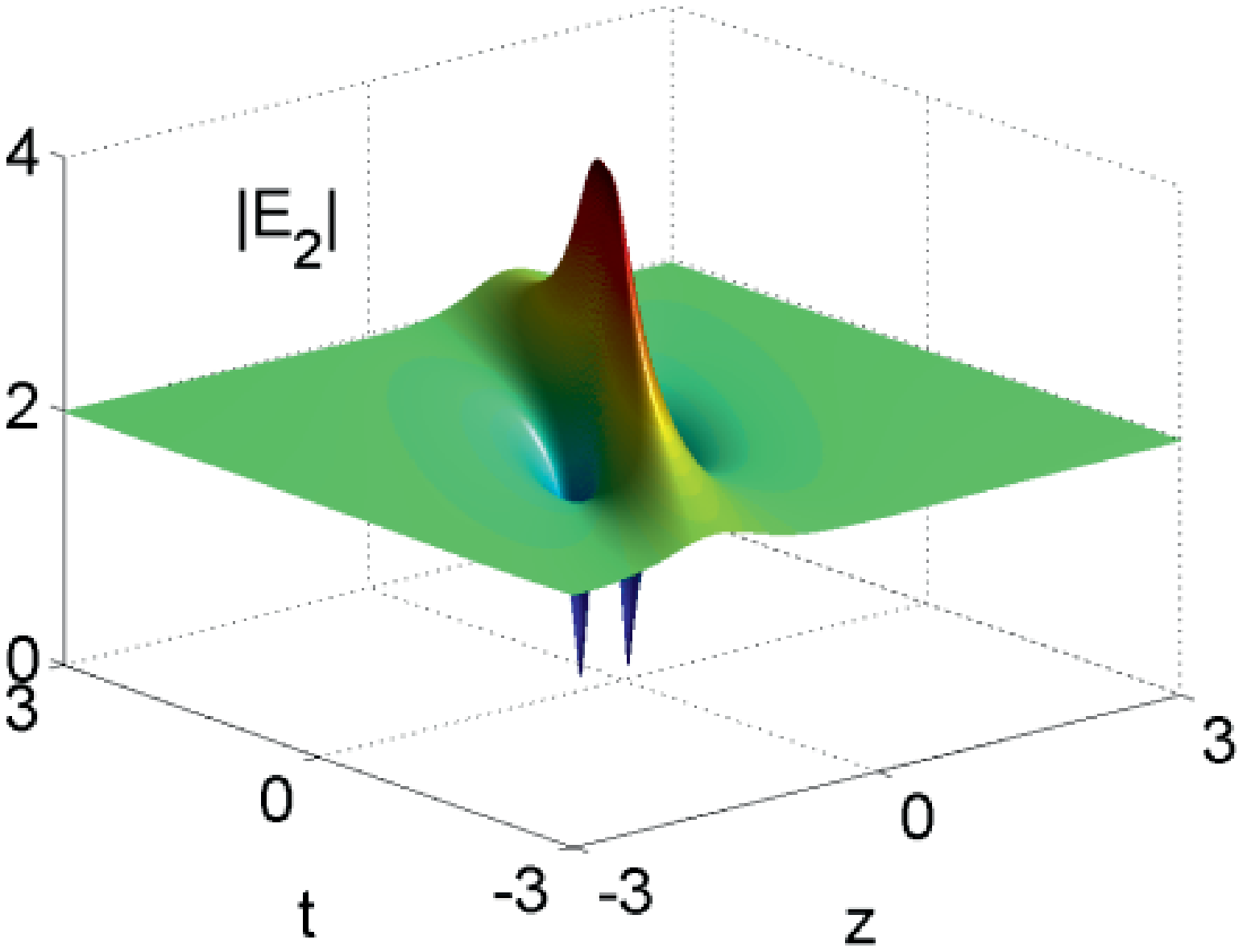}
\includegraphics[width=6cm]{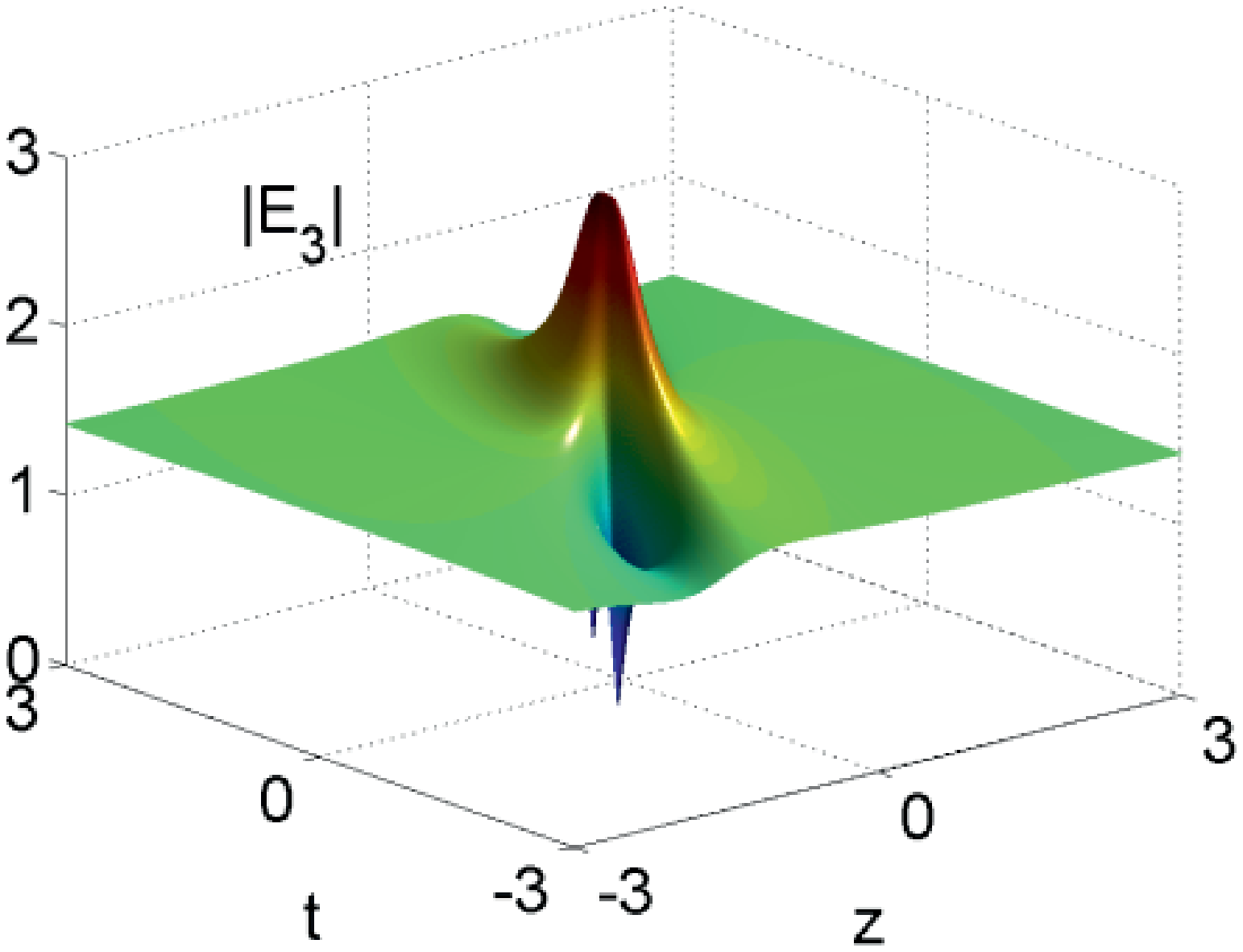}
    \end{center}
     \caption{Vector rogue waves envelope distributions $|E_1|$, $|E_2|$
     and $|E_3|$ of (\ref{pere}). Here, $V_1=1, V_2=0.5, q=1, \gamma_1=1, \gamma_2=1, \gamma_3=0$.
    } \label{fig_peregrine}
\end{figure}

However not all these parameters are essential, since some of them can be fixed without loosing generality by using appropriate symmetries of the TWRI equations (\ref{3wri}). The parameter $q$ merely rescales the wave amplitudes, and the coordinates $z$ and $t$. Thus one 
can set $q=1$. 

Also the three remaining parameters $\gamma_1, \gamma_2, \gamma_3$ are not all essential as one 
(non vanishing) of them can be given the unit value. Moreover it can be shown that if $\gamma_2 = \gamma_3 =0$ 
then the solution (\ref{pere}) represents plane wave backgrounds with no interest. 
Otherwise, if $\gamma_3 =0$ then the parameter $\gamma_1$ can be made to vanish by using 
translation invariance, while, by the same argument, one can set $\gamma_2=0$ if $\gamma_3\neq 0$, while $\gamma_1$ remains instead an essential parameter.
Despite this simplification, we choose to keep all three parameters $\gamma_1, \gamma_2, \gamma_3$ and 
to play with them to better display a few aspects of the many properties of this family of solutions (\ref{pere}).  
  

%

\begin{figure}[h]
\begin{center}
\includegraphics[width=6cm]{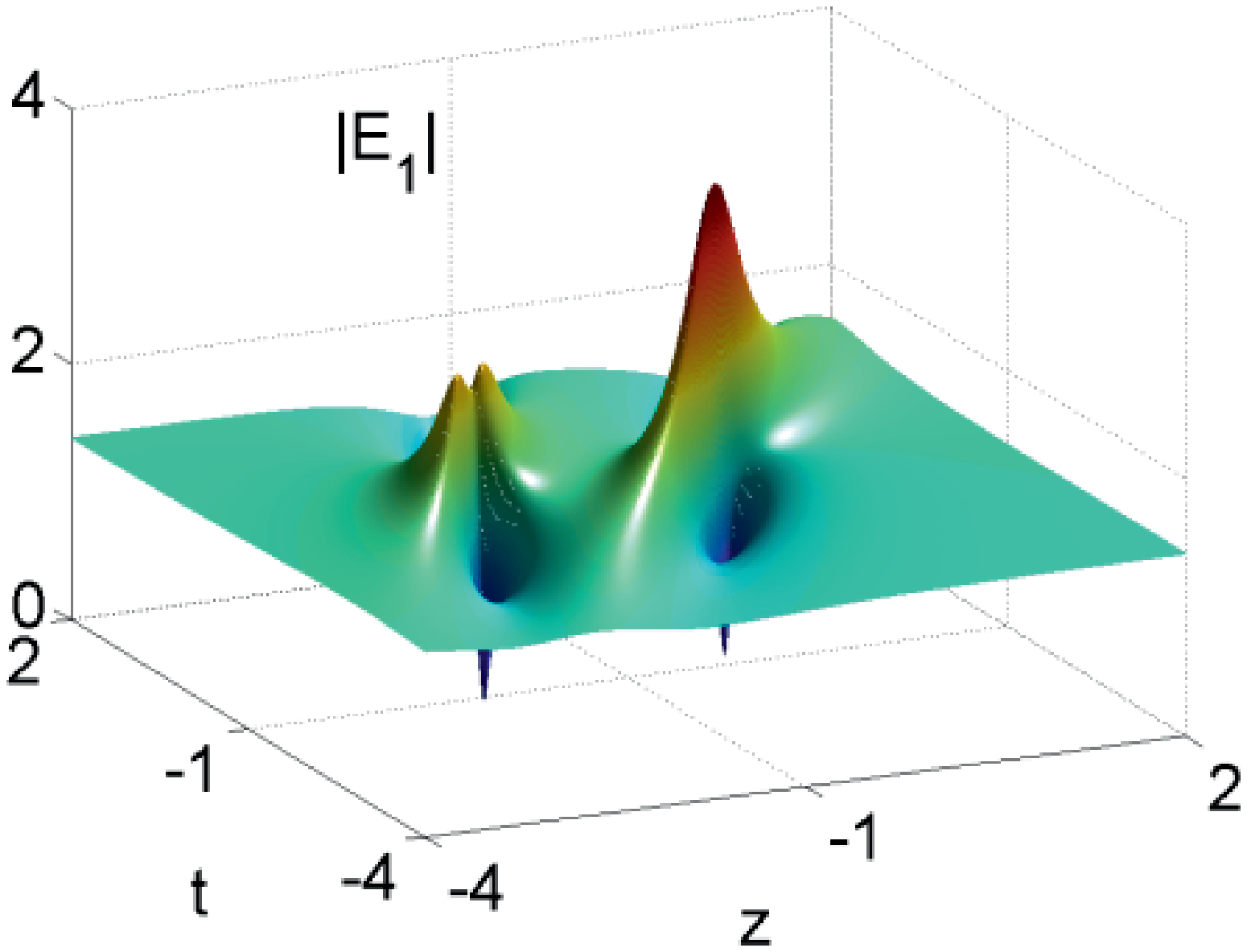}
\includegraphics[width=6cm]{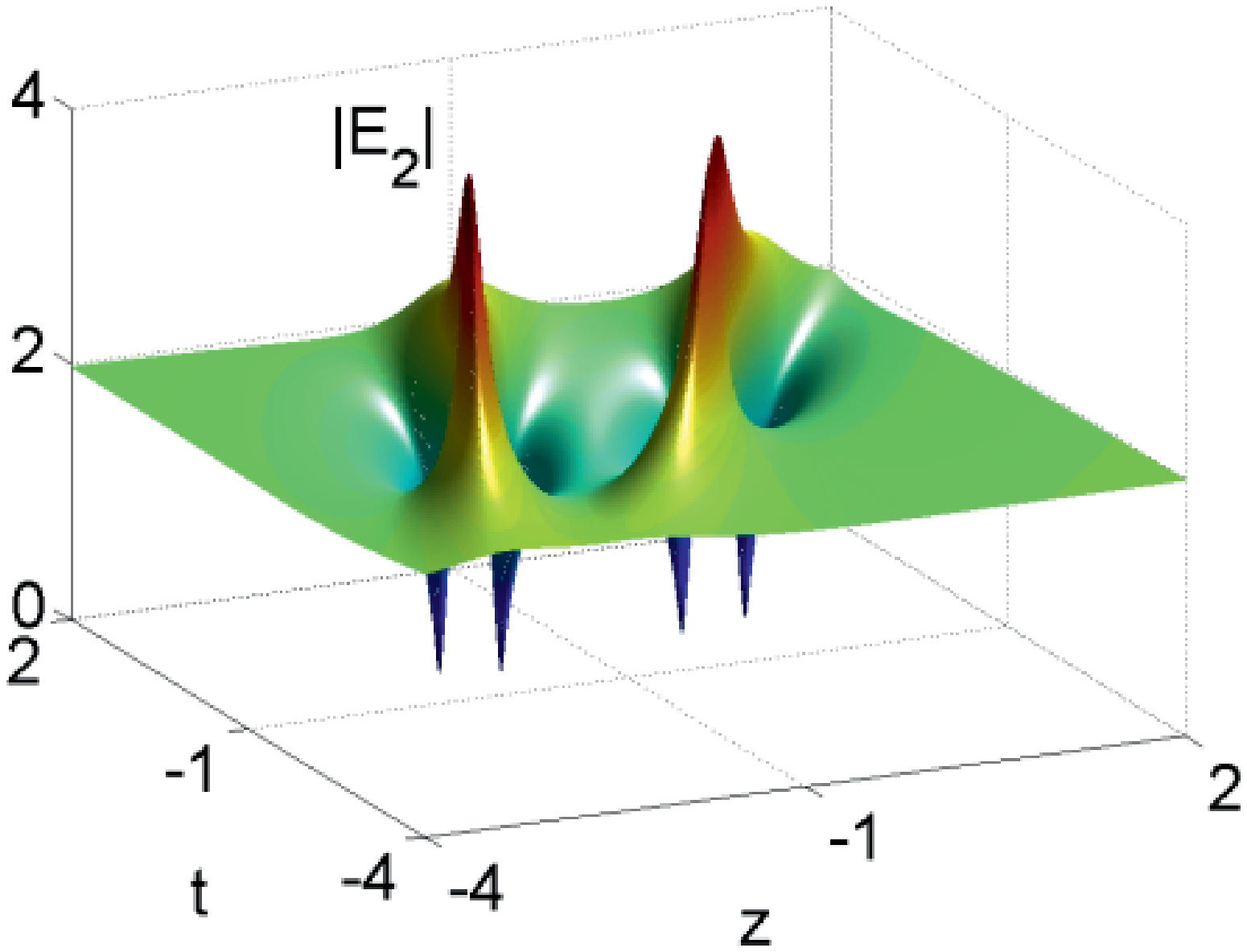}
\includegraphics[width=6cm]{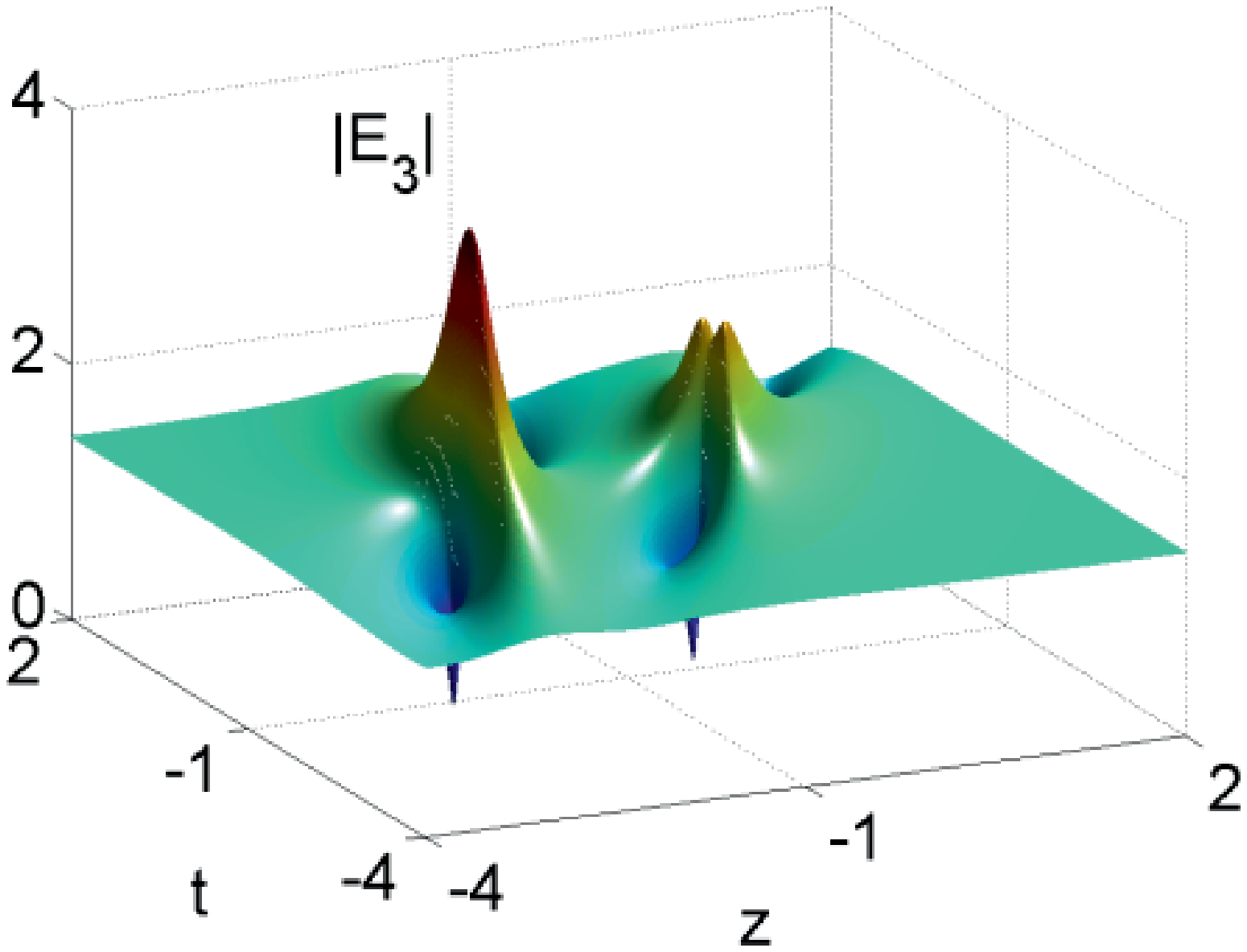}
    \end{center}
     \caption{Vector rogue waves envelope distributions $|E_1|$, $|E_2|$
     and $|E_3|$ of (\ref{pere}). Here, $V_1=1, V_2=0.5, q=1, \gamma_1=2, \gamma_2=7, \gamma_3=1.5+i$.
    } \label{fig_peregrine2}
\end{figure}

\begin{figure}[ht]
\begin{center}
\includegraphics[width=6cm]{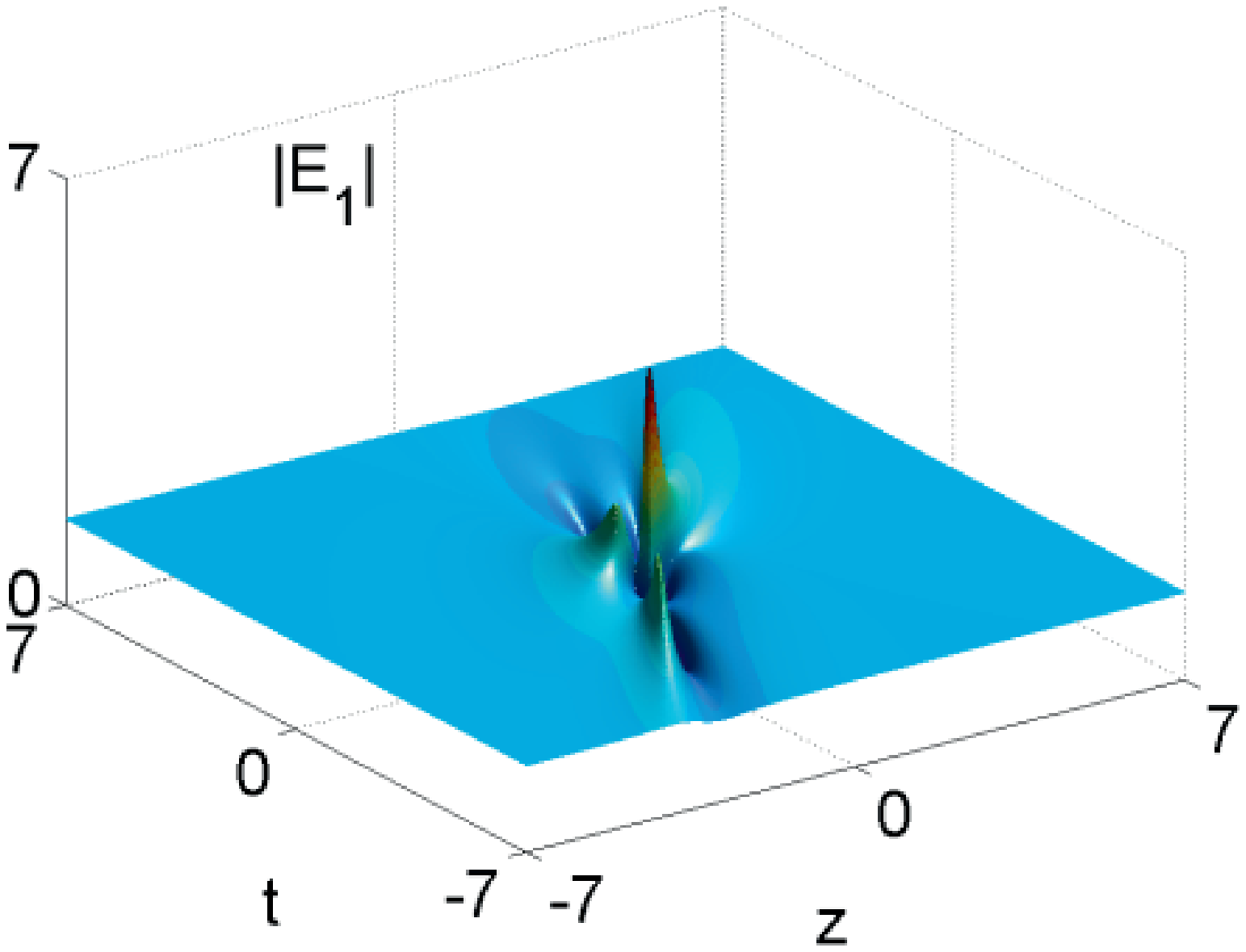}
\includegraphics[width=6cm]{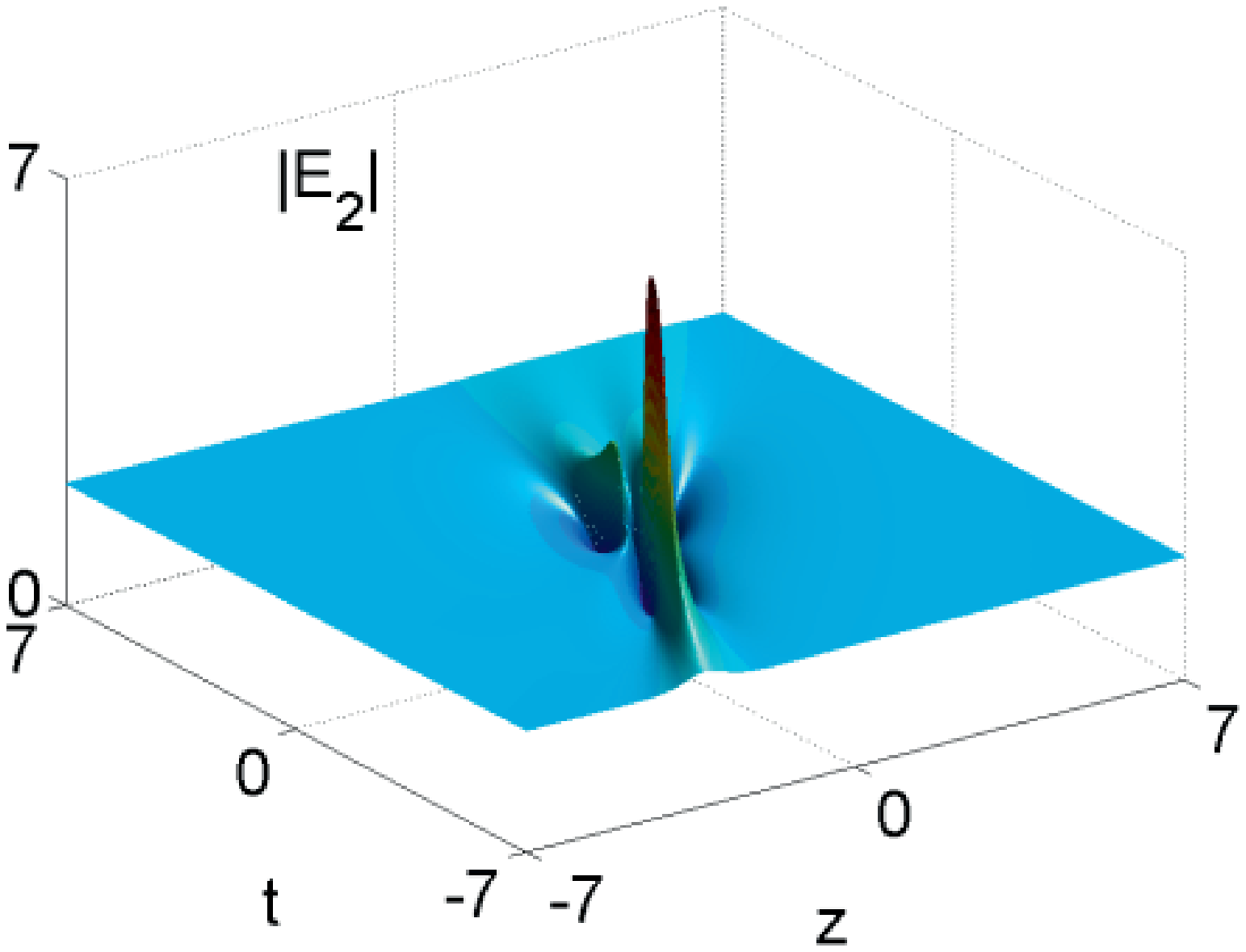}
\includegraphics[width=6cm]{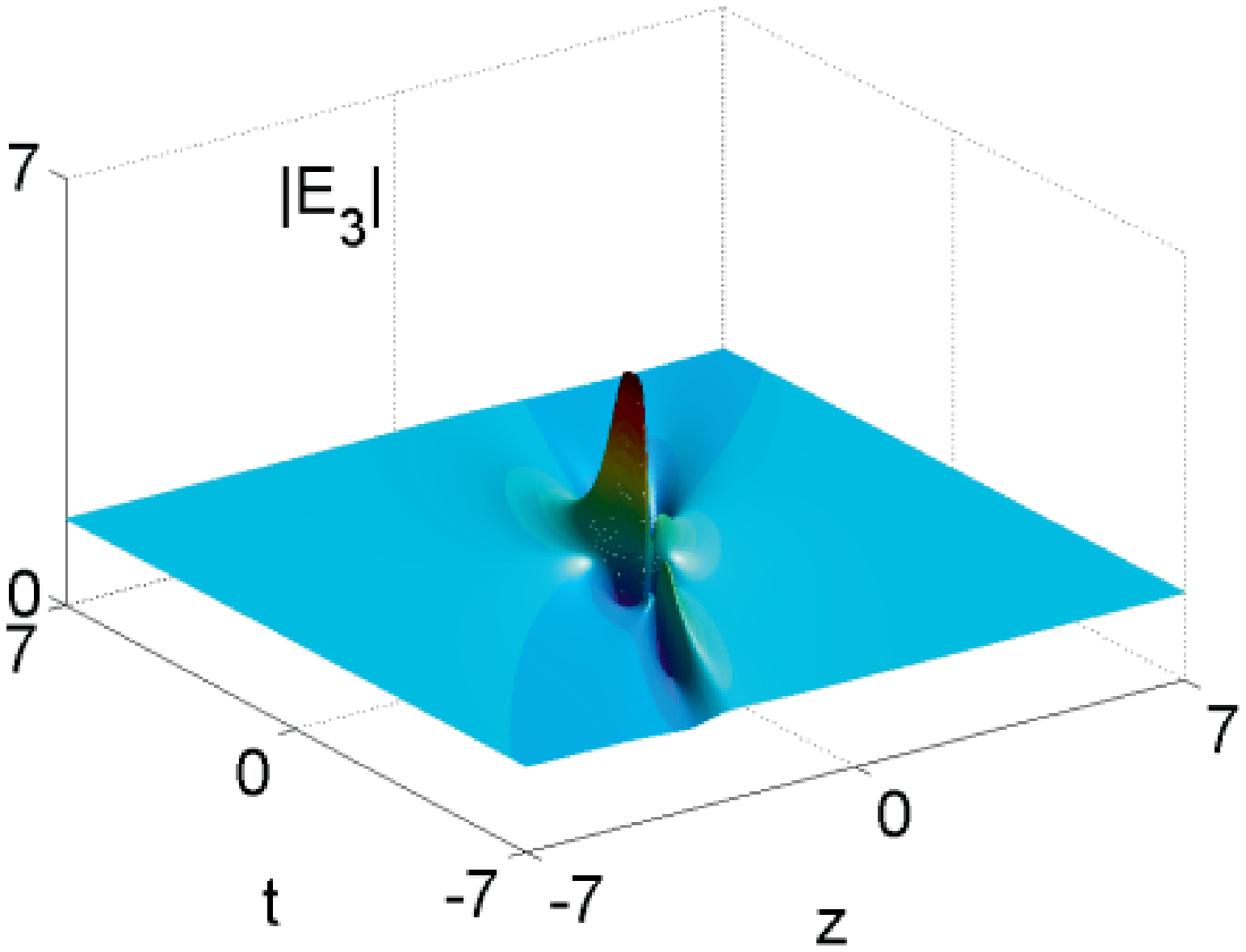}
    \end{center}
     \caption{Vector rogue waves envelope distributions $|E_1|$, $|E_2|$
     and $|E_3|$ of (\ref{pere}). Here, $V_1=1, V_2=0.5, q=1, \gamma_1=1, \gamma_2=0.1, \gamma_3=i$.
    } \label{fig_peregrine3}
\end{figure}

In Fig. \ref{fig_peregrine}, we first show the case $ \gamma_2 \neq 0, \gamma_3=0$. The parameter $\gamma_1$ is so chosen as to put the peak at the origin of the $(z,t)$ plane. As expected, the expression (\ref{pere})
describes amplitude peaks which are localized in both $z$ and $t$. Interestingly, 
each component $|E_n|$ looks like a rogue wave whose maximum height is  twice the background intensity while its minimum is zero; its eye-shaped distribution density shows one hump and two valleys. We also note that, as for the Peregrine soliton, 
the rational expression (\ref{pere}) is the ratio of two polynomials of second degree in the coordinates $z,t$.

In the case $ \gamma_3 \neq 0$, the expression (\ref{pere}) may
describe amplitudes with multiple peaks localized in $z$ and $t$. 
Figure \ref{fig_peregrine2} shows two rogue waves with different structures in each one of the three components $|E_n|$. 
Figure \ref{fig_peregrine2} shows, in $E_1$ component, a bright rogue wave with 
an eye-shaped distribution (a hump and two valley), together with a wave with 
a four-petaled distribution (two humps and two valleys around a center, and the center 
value is almost equal to that of the background). The four-petaled wave in $E_1$ component 
corresponds to eye-shaped rogue waves in $E_2$ and $E_3$ components. 
By decreasing the value $|\gamma_3/ \gamma_2|$, in each component these rogue waves separate. By
increasing $|\gamma_3/ \gamma_2|$, these rogue waves instead merge, giving birth to  higher-amplitude vector rogue wave solutions (see Fig. \ref{fig_peregrine3}). 
The maximum value of the humps is more than 3 times the plane wave's background
for some components, and the minimum value of the valleys is zero.

\begin{figure}[ht]
\begin{center}
\includegraphics[width=4cm]{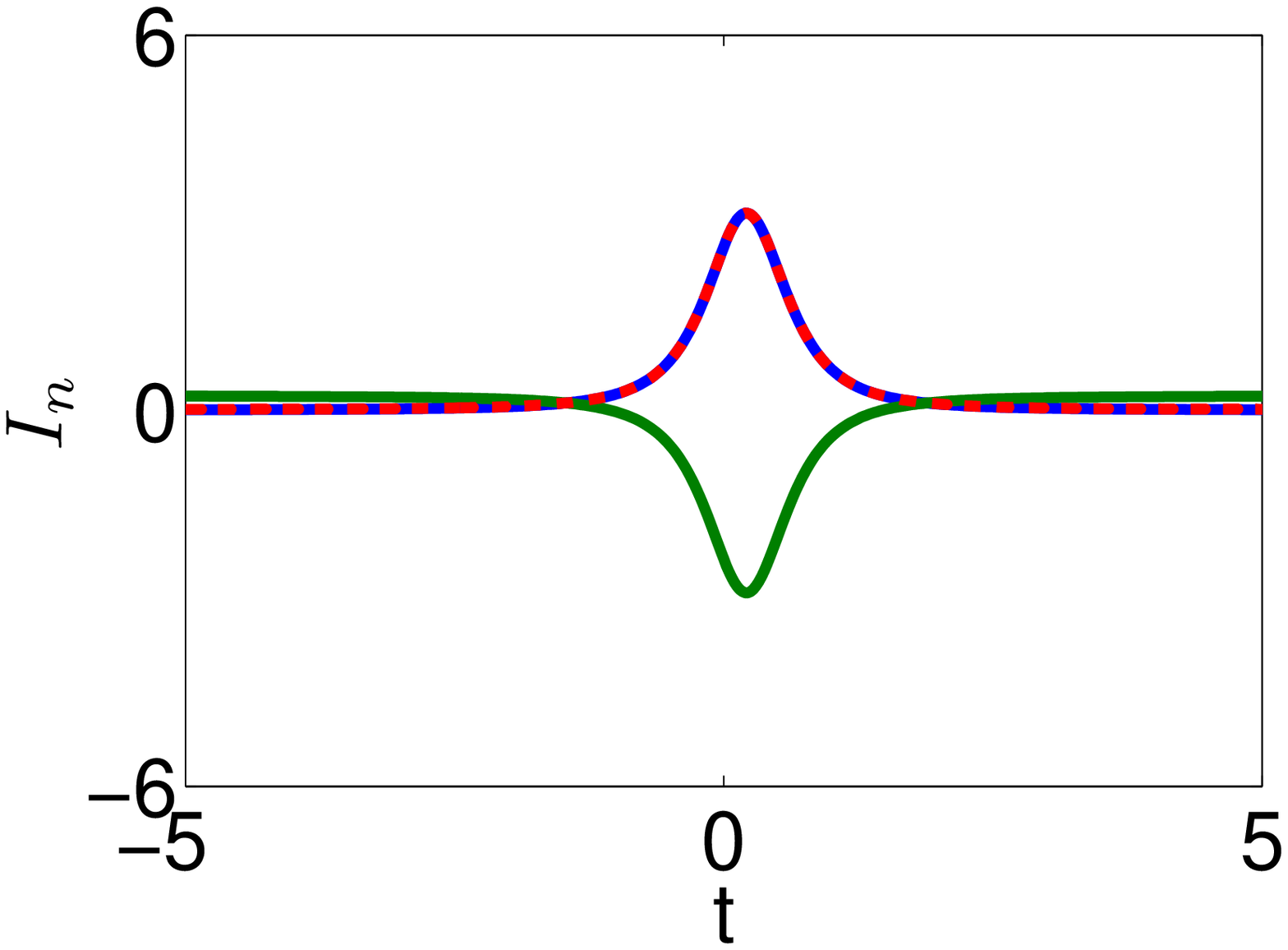},\includegraphics[width=4cm]{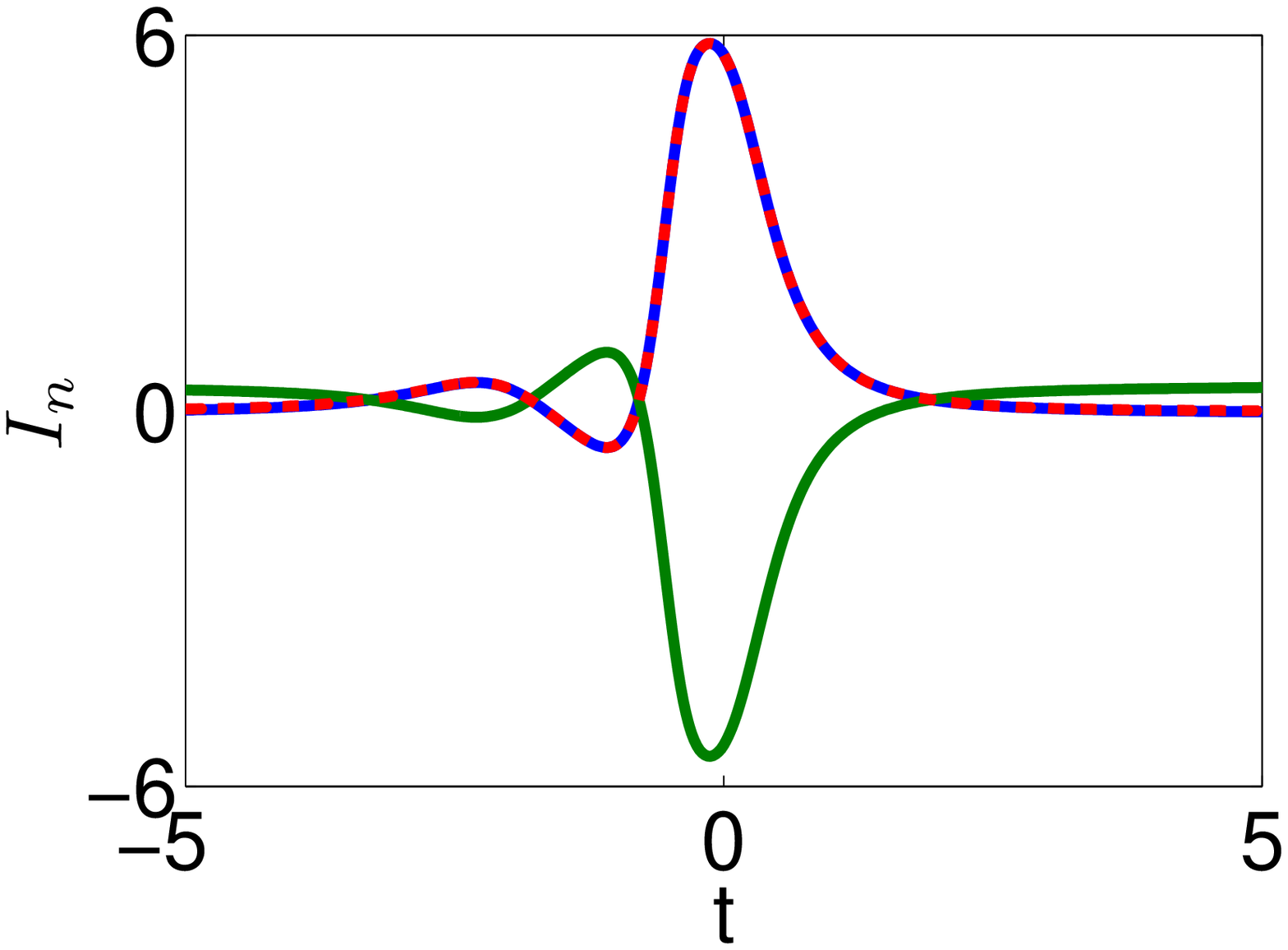}
    \end{center}
     \caption{Effective energy evolution $\overline I_{1}$ (blue line), $\overline I_{2}$ (green line) and
$\overline I_{3}$ (red line) versus $t$, for the case reported in Fig. \ref{fig_peregrine} (left) and
Fig. \ref{fig_peregrine3} (right).} \label{energie}
\end{figure}

Notice that effective energy exchanges take place between waves $E_1$, $E_2$
and $E_3$ in TWRI during their interaction.  
Figure \ref{energie} reports a typical evolution of the effective energy $\overline I_{1}$, $\overline I_{2}$ and
$\overline I_{3}$ versus $t$.
Effective energies $\overline I_{1}$, $\overline I_{2}$ and
$\overline I_{3}$ are obtained according to the prescription:
\begin{eqnarray}\label{rrrrr}
\nonumber \overline I_{n}&=&\frac{1}{2} \int (|E_n|^2-|E_{n0}|^2)dz,
\end{eqnarray}
where $E_{n0}=\lim_{z\rightarrow \infty}E_n$, $n=1,2,3$, is the plane wave background. 
The energy transfer between the waves can enhance the peak 
amplitude in some of the wave components.
This wave behavior is completely different from what happens in 
coupled NLSEs, where energy exchanges are forbidden.

Let us briefly discuss the experimental condition in nonlinear optics for the observation of TWRI rogue waves. In fact, nonlinear optics has been recently seen as
a fertile, reproducible and safe ground to experimentally develop the knowledge of 
rogue waves \cite{erkin2009,akhmediev2010sp,onorato13,kibler10}.
One may consider a TWRI optical spatial non-collinear scheme with type II second-harmonic generation 
in a $3$-cm long birefringent KTP crystal (f.i., see the experimental set-up of Ref. \cite{baronio10}). 
Spatial diffraction-less $6$-mm waist beams, mimicking quasi-plane waves, at $1064$ nm (o-wave, and e-wave)
and at $532$ nm (e-wave) would lead to TWRI modulational instability evidence and rogue wave dynamics with peak
field intensities of tens of MW/cm$^2$.

As a final remark, for the way of computing the expression (\ref{pere}), we limit ourselves to notice that this is 
based on the Darboux technique applied to the Lax pair associated to the TWRI equations. This method is 
well known and does not need to be detailed here to any extent. The relevant literature is rather vast 
and we refer to \cite{DL09} for the formalism we have adopted and to \cite{baronio12,DL13} for the basic 
arguments to follow for the construction of rational solutions.

\textit{Conclusions.}\label{sec3T}
We have reported the explicit analytic expression of solutions of the equations describing the resonant interaction of three waves. These solutions have the important property of describing rogue wave events.
Several articles have been recently devoted to rogue waves as rational solutions of 
multi-component systems of coupled wave equations:  VNLS equations \cite{baronio12,onorato12g,liu2013,zhai2013},
 Davey-Stewartson equation \cite{ohta12} and coupled Hirota 
systems \cite{chen13}.

%
The present step in this direction, dealing with rogue wave solutions of the TWRI equations which represents a fundamental and universal model for the 
description of strong resonant interactions, seems to be a  crucial stride  to controlling and forecasting  extreme-wave dynamics in multi-component wave systems, with a broad variety of applications.

The present research was supported by the Italian Ministry of University and Research (MIUR, Project Nb.2009P3K72Z),
by the  Italian Institute for Nuclear Physics (INFN Project Nb. RM41) and by the Netherlands 
Organisation for Scientific Research (NWO, Grant 639.031.622).

\end{document}